\documentclass{article}
\usepackage[a4paper, left=2.5cm, right=2.5cm, top=2.5cm, bottom=2.5cm]{geometry}
\usepackage[utf8]{inputenc}
\usepackage[T1]{fontenc} 
\usepackage{xcolor}
\usepackage{array}
\usepackage[table]{xcolor}
\usepackage[nottoc,numbib]{tocbibind}
\usepackage{bm}
\usepackage{algorithm}
\usepackage{algorithmic}

\usepackage{booktabs}
\usepackage{makecell}

\usepackage{graphicx} 
\usepackage{float}
\usepackage{amsthm} 
\usepackage{amssymb} 
\usepackage{amsmath}
\usepackage{authblk}  
\usepackage{hyperref}
\usepackage{subcaption}
\usepackage[font=small]{caption}
\usepackage[authoryear,round]{natbib}

\definecolor{lightgreen}{RGB}{200,255,200}

\newcommand{\bigbinom}[2]{\left(\genfrac{}{}{0pt}{}{#1}{#2}\right)}

\newenvironment{keywords}{
  \begin{quote}
  \textit{Keywords}:
}{
  \end{quote}
}

\usepackage{titletoc}
\titlecontents{section}%
  [1.5em]
  {\normalsize}
  {\contentslabel{2.3em}}
  {}
  {\titlerule*[0.5pc]{.}\contentspage}
\newtheorem{tw}{Theorem} 
\newtheorem{fakt}{Fact}
\newtheorem{obs}{Remark}
\newtheorem{prop}{Proposition}
\newtheorem{wniosek}{Corollary}

\title{The post-hoc test for local dependence}

\author[1]{Bogdan Ćmiel}
\author[2]{Bartłomiej Gibas}

\affil[1]{AGH University of Krakow, Faculty of Applied Mathematics,\newline Al. Mickiewicza 30, 30-059 Kraków, Poland}
\affil[2]{AGH University of Krakow, Faculty of Computer Science, Electronics \newline and Telecommunications, Al. Mickiewicza 30, 30-059 Kraków, Poland}

\date{}

\begin{document}
\maketitle
\begin{abstract}
The concept of independence plays a crucial role in probability theory and has been the subject of extensive research in recent years. Numerous approaches have been proposed to test for independence; however, most of them address the problem only at a global level. From a practical perspective, it is important not only to determine whether the data are dependent but also to identify where this dependence occurs and how strong it is. The graphical presentation of results is another essential aspect that should not be neglected, as it considerably enhances interpretability.

The main objective of this work is to propose a solution that considers these aspects simultaneously. Relying on copula-based results, we introduce a novel method for testing global and local statistical independence using the quantile dependence function. Rather than assessing whether the value of the test statistic exceeds a single critical threshold and subsequently deciding whether to reject the independence hypothesis, we introduce so-called critical surfaces that guaranty a locally equal probability of exceeding them under independence. This approach enables a detailed examination of local discrepancies and an assessment of their statistical significance while preserving the overall significance level of the test. 

\end{abstract}


\begingroup
\renewcommand\thefootnote{}
\footnotetext{\textit{Email addresses:} \texttt{cmielbog@gmail.com} (Bogdan Ćmiel), \texttt{bartekgibas13@gmail.com} (Bartłomiej Gibas)}
\endgroup

\newpage
\tableofcontents
\newpage
\section{Introduction}
\hspace*{\parindent}The concept of independence between two or more random variables is one of the most important topics in statistics. Despite its simplicity, it is a powerful tool and often the starting point for advanced data analysis and building machine or deep learning models. These areas have become increasingly popular in recent years, which makes it necessary to develop existing approaches or introduce new ones. Therefore, particular attention should be paid to methods for testing this independence effectively. 

The problem of investigating independence can be considered both at the global and local levels. The first approach answers the question of whether the data are dependent or not. There are numerous studies on this topic, including methods using kernel function (\cite{Gretton}), empirical copula (\cite{Genest04}), weighted empirical process (\cite{Deheuvels}, \cite{Berghaus}, \cite{Cmiel}), local Gaussian correlation (\cite{JonesM.C.2003DmLd}), sign covariance (\cite{Bergsma}), point process (\cite{Dvorak}) and others. 

We focus primarily on the local approach. An early treatment of this problem can be found in the work of  \cite{HollandPaulW.1986RDFC}, who discuss the theoretical foundations for identifying local discrepancies from independence in the bivariate case by investigating the second derivatives of the log-transformed joint distribution. This work is further developed by \cite{JonesM.C.2003DmLd}, where estimation procedures based on kernel methods are proposed. The authors also introduce a graphical representation of the results using dependence maps, which reveal patterns that indicate the presence of a relationship between variables. Similar findings are reported by \cite{Berentsen}, who approximate locally the unknown density using a Gaussian distribution and then compare the value of the normal correlation with zero. Another approach to detecting local dependence is provided by scanning or binning methods. These procedures partition the sample space into bins and analyse the dependence structure in local parts of the space. \cite{MaLi2019FESf} connect Fisher's exact test and multi-scale scanning. In other words, the sample space is examined in multiple resolutions, and Fisher’s exact test is applied within the resulting bins to identify local departures from independence. A similar idea but built on binary expansion can be found in the work of \cite{ZhangKai2019BoI}.

Notable results are presented in \cite{cmiel2024detectingdependencestructurevisualization}. They define a grid and construct a global test of independence based on the quantile dependence function (some weighted empirical process). In addition, they propose a local testing procedure in which the lower and upper barriers are defined for selected parts of the grid. Crossing these barriers leads to the rejection of the independence hypothesis in the corresponding region. As a result, acceptance regions can be identified. However, this technique controls the significance level for individual local tests within a particular area but does not guarantee control of the overall significance level. If the global test is performed at the level $\alpha$, then applying the same level to all local tests does not preserve the overall significance level. This occurs due to the multiple testing problem, which requires the use of appropriate correction procedures. Even when such corrections are applied, the overall power of such procedure may decrease substantially. A second important aspect is the fact that, for a sufficiently large sample size within a single acceptance region, both barriers may be crossed by different points in this area, which may lead to ambiguity.

Our approach is based on the results of \cite{cmiel2024detectingdependencestructurevisualization}. Similarly, we construct a test based on the quantile dependence function, but we perform local tests with appropriate significance levels  to ensure that the overall significance level (family error rate) remains at the chosen level $\alpha$. Every single local test has its own acceptance region. The lower and upper bounds of those regions create our lower and upper critical surfaces for the quantile dependence function estimator. In our method, the test statistic may exceed only one surface within a particular area, making it possible to clearly establish the nature of the relationship. Since our approach is distribution-free (rank-based), we are able to construct a procedure that is not conservative and, as simulations show, almost as powerful as a global test from \cite{cmiel2024detectingdependencestructurevisualization}. In general, the proposed procedure allows for testing global independence and identifying regions where departures from independence are significant while controlling the overall significance level. In some sense, the interpretation of our test is similar to post-hoc analysis in many group comparisons, for example, Tukey's HSD in ANOVA. One can test for many pairs of quantiles if there are significant deviations from hypothesis (independence).

The structure of the paper is as follows. In Section 2, we present a functional measure of local dependence and the corresponding critical surfaces used to construct the test. In Section 3, we introduce tests for global and local dependence and formulate a theorem that establishes the consistency of this test. In Section 4, we show examples of applying our test and visualising local dependence for certain real-world data. We also present simulation results on the test’s power in a broad class of alternatives. In Section 5, we discuss the implementation details. In Section 6, we provide the proof of the main theorem. Supplementary materials includes an additional example and a comment on convergence of local significance.

\section{Quantile dependence function and critical surfaces}

\hspace*{\parindent} Let $X$ and $Y$ be a pair of random variables with a joint distribution $H$ and continuous marginal distributions $F$ and $G$, respectively. By Sklar's theorem  (\cite{alma991053257912808832}), there exists a uniquely determined copula $C$ defined on $[0,1]\times[0,1]$ such that $C(u,v) = H(F^{-1}(u), G^{-1}(v)),$ where $F^{-1}(y) = \inf\{x\in\mathbb{R}: F(x)\geq y\}$ is the quantile function evaluated at $y\in[0,1].$ Similarly $G^{-1}.$ We consider the copula-based measure of dependence  
\begin{align}\nonumber
    q(u,v) = \frac{C(u,v)-uv}{\sqrt{u(1-u)v(1-v)}}, \hspace{0.5cm} \text{for} \hspace{0.5cm} (u,v)\in(0,1)^2,
\end{align}
which is called the quantile dependence function. For the properties of $q$, see  \cite{ledwina2014dependencefunctionbivariatecdfs} and \cite{Ledwina2015}.
\begin{obs}
If $q(u,v)>0$ for any $(u,v)\in(0,1)^2,$ we say that $(X,Y)$ are locally, for the pair of quantiles $(x_u,y_v)$, positive dependent. We define negative dependence analogously. The random variables $X$ and $Y$ are independent if and only if $q\equiv 0$. For the interpretation of local positive and local negative dependency, see Proposition 1 in \cite{cmiel2024detectingdependencestructurevisualization}. 
\end{obs}
Based on $(X_1,Y_1),\ldots,(X_n,Y_n)$ i.i.d. from the distribution $H$, the natural estimator of $q$ is   
\[
q_n(u,v) = \frac{C_n(u,v)-uv}{\sqrt{uv(1-u)(1-v)}},
\]
where $C_n(u,v) = 1/n\sum_{i=1}^{n}1(R_i/n\leq u, S_i/n\leq v)$ is a copula estimator at $(u,v)$. Here, $R_i$ and $S_i$ denote the ranks of $X_i$ and $Y_i$ within the marginal samples $X_1,\ldots,X_n$ and $ Y_1,\ldots,Y_n$, respectively. Note that the continuity of marginal distributions implies that, with probability one, there are no ties. 

The problem of testing $H_0:$ $X$ and $Y$ are independent against $H_1:$ $X$ and $Y$ are dependent is equivalent to
$$H_0:\ \forall\ (u,v)\in(0,1)^2 \ \ q(u,v)=0, \ \ \ \ \text{against}\ \ \ \ \ H_1:\ \exists\ (u,v)\in(0,1)^2 \ \ q(u,v)\neq0.$$

A straightforward way to test independence using the quantile dependence function is to reject $H_0$ whenever $q_n$ becomes large enough, i.e., when its norm exceeds the critical threshold $c_{\alpha,n}$ set for a chosen significance level $\alpha$. In such an approach, for $||\cdot||_{\infty}$, we have 
\begin{align}\nonumber
    P_{H_0}\Big(\exists_{(u,v)\in(0,1)^2} \ \left|q_n(u,v)\right|> c_{\alpha,n} \Big) \leq \alpha. \label{istotność_alpha}
\end{align}
Observe that the inner inequality can be rewritten as the disjunction of two conditions
\[
P_{H_0}\Bigg(\exists_{(u,v)\in(0,1)^2}: \quad \Big\{-c_{\alpha,n}>q_n(u,v)\Big\} \quad \cup \quad  \Big\{q_n(u,v)>c_{\alpha,n}\Big\}\Bigg) \leq \alpha.
\]
In our method, we seek two real-valued functions satisfying certain assumptions rather than a single constant, allowing their values to vary from point to point. Based on this idea, we can modify the simple approach to 
\begin{align}
P_{H_0}\Bigg(\exists_{(u,v)\in(0,1)^2}: \quad \Big\{l_{\alpha,n}(u,v)>q_n(u,v)\Big\} \quad \cup \quad  \Big\{q_n(u,v)>u_{\alpha,n}(u,v)\Big\}\Bigg) \leq \alpha. \label{alpha}
\end{align}
Any pair of functions $l_{\alpha,n}$ and $u_{\alpha,n}$ that satisfy the above condition we can call \textit{the critical surfaces} (lower and upper). To treat positive and negative dependencies equally, we want the probability of exceeding the upper and lower critical surfaces to be as equal as possible at each point. We can achieve this by keeping every test for each point $(u,v)$ at the same significance level $\eta$ smaller than $\alpha$. Let us take $L_{\eta}$, $U_{\eta}$ so that for all $(u,v)\in(0,1)^2$
$$P_{H_0}\Big(q_n(u,v) < L_{\eta}(u,v)\Big)\leq\eta/2, \ \ P_{H_0}\Big(q_n(u,v) \leq L_{\eta}(u,v)\Big)\geq\eta/2,$$
$$P_{H_0}\Big(q_n(u,v) > U_{\eta}(u,v)\Big)\leq\eta/2,\ \  P_{H_0}\Big(q_n(u,v) \geq U_{\eta}(u,v)\Big)\geq\eta/2.$$
Notice that $L_{\eta}(u,v)$, $U_{\eta}(u,v)$ are $q_n(u,v)$ quantiles of order $\eta/2$, $1-\eta/2$, respectively. Now we take the maximal $\eta$ that keeps the significance level $\alpha$, i.e.,
\begin{align}
    \eta_n(\alpha) = \sup\Bigg\{\eta\in[0,\alpha]: P_{H_0}\Bigg(\exists_{(u,v)\in(0,1)^2}: \ \Big\{L_{\eta}(u,v)>q_n(u,v)\Big\} \ \cup \  \Big\{q_n(u,v)>U_{\eta}(u,v)\Big\}\Bigg) \leq \alpha \Bigg\}. \nonumber  
\end{align}
We call $\eta_n(\alpha)$ \textit{the local significance level}. The critical surfaces $L_{\eta_n(\alpha)}$ and $U_{\eta_n(\alpha)}$ are uniquely determined.
To calculate  $(\eta_n(\alpha), L_{\eta_n(\alpha)}, U_{\eta_n(\alpha)})$ we do not need to inspect all $(u,v)\in(0,1)^2$ points. Recall that the copula's estimator $C_n$ for each point $(u,v)\in \big[i/n,(i+1)/n\big)\times\big[j/n,(j+1)/n\big)$ for some $i,j\in\{0,\ldots,n-1\},$ is constant. One can easily see that for $i=0$ or $j=0$ the estimator $C_n$ is equal to 0 with probability one. 
Therefore, the distribution of $q_n$ is determined by $C_n$ on the grid $\{1,...,n-1\}^2$. Due to this fact, we only need to calculate a finite number of quantiles $L_{\eta}(i/n,j/n)$ and $U_{\eta}(i/n,j/n)$ for all $(i,j)\in\{1,\dots,n-1\}^2$
and
\begin{align}
\eta_n(\alpha) = {} & \sup\Bigg\{\eta\in[0,\alpha]: \label{kryt_lh} \\
& P_{H_0}\Bigg(\exists_{(i,j)\in\{1,\dots,n-1\}^2}:
\left\{L_{\eta}\left(\frac{i}{n},\frac{j}{n}\right)>q_n\left(\frac{i}{n},\frac{j}{n}\right)\right\}
\cup
\left\{q_n\left(\frac{i}{n},\frac{j}{n}\right)>U_{\eta}\left(\frac{i}{n},\frac{j}{n}\right)\right\}\Bigg)
\leq \alpha \Bigg\}. \nonumber
\end{align}
Using the Bonferroni correction for multiple testing problems, one can easily prove the following proposition.
\begin{prop}\label{prop1}
    For any fixed sample size $n\geq2$ and a significance level $\alpha$ it holds that 
    $$\frac{\alpha}{(n-1)^2} \leq \eta_n(\alpha) \leq \alpha.$$
\end{prop}
The true value of $\eta_n(\alpha)$ is difficult to calculate, but since the distribution of $q_n$ does not depend on the marginal distributions ($F$ and $G$) under $H_0$, it is possible to obtain $\eta_n(\alpha)$ using the Monte Carlo method.
\begin{obs}
Using \eqref{kryt_lh} we obtain $L_{\eta_n(\alpha)}(i/n,j/n)$ and $U_{\eta_n(\alpha)}(i/n,j/n)$ for all $(i,j)\in\{1,\dots,n-1\}^2$. For any $(u,v)\in[i/n,(i+1)/n)\times[j/n,(j+1)/n)$ the values of $L_{\eta_n(\alpha)}(u,v)$ and $U_{\eta_n(\alpha)}(u,v)$ are determined by $L_{\eta_n(\alpha)}(i/n,j/n)$ and $U_{\eta_n(\alpha)}(i/n,j/n)$, i.e., we have
$$L_{\eta_n(\alpha)}(u,v)=\frac{L_{\eta_n(\alpha)}(i/n,j/n)\sqrt{\frac{i}{n}\frac{j}{n}(1-\frac{i}{n})(1-\frac{j}{n})}+\frac{i}{n}\frac{j}{n}-uv}{\sqrt{uv(1-u)(1-v)}},$$
$$U_{\eta_n(\alpha)}(u,v)=\frac{U_{\eta_n(\alpha)}(i/n,j/n)\sqrt{\frac{i}{n}\frac{j}{n}(1-\frac{i}{n})(1-\frac{j}{n})}+\frac{i}{n}\frac{j}{n}-uv}{\sqrt{uv(1-u)(1-v)}}.$$
\end{obs}

\section{The test}
\begin{figure}[H]
  \centering
  \includegraphics[width=0.9\textwidth]{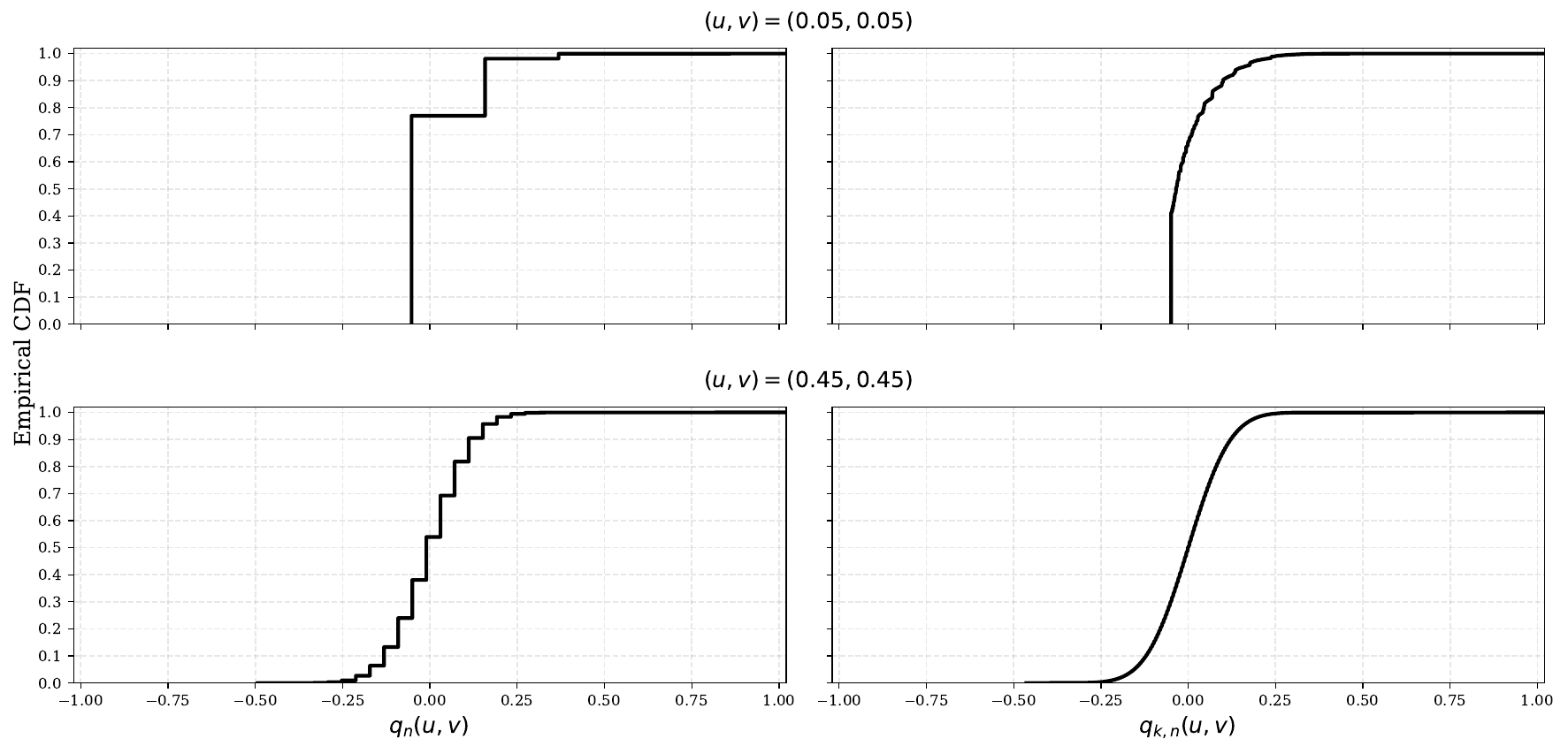}
  \caption{Empirical cumulative distribution functions of $10^6$ Monte Carlo realizations for $q_n$ and $q_{k,n}$ ($k=11$) based on 100 samples generated from uniform distribution on the interval $(0,1)$ in points: $(0.05, 0.05)$ and $(0.45, 0.45).$ The first column presents the values of the estimator before smoothing, whereas the second column shows the corresponding values after the smoothing procedure.}
  \label{fig:histograms_bar_q}
\end{figure}

\begin{figure}[h]
    \centering
    \scalebox{0.9}{%
        \begin{minipage}{\linewidth}
            \centering
            \begin{subfigure}{\linewidth}
                \centering
                \includegraphics[width=0.75\linewidth]{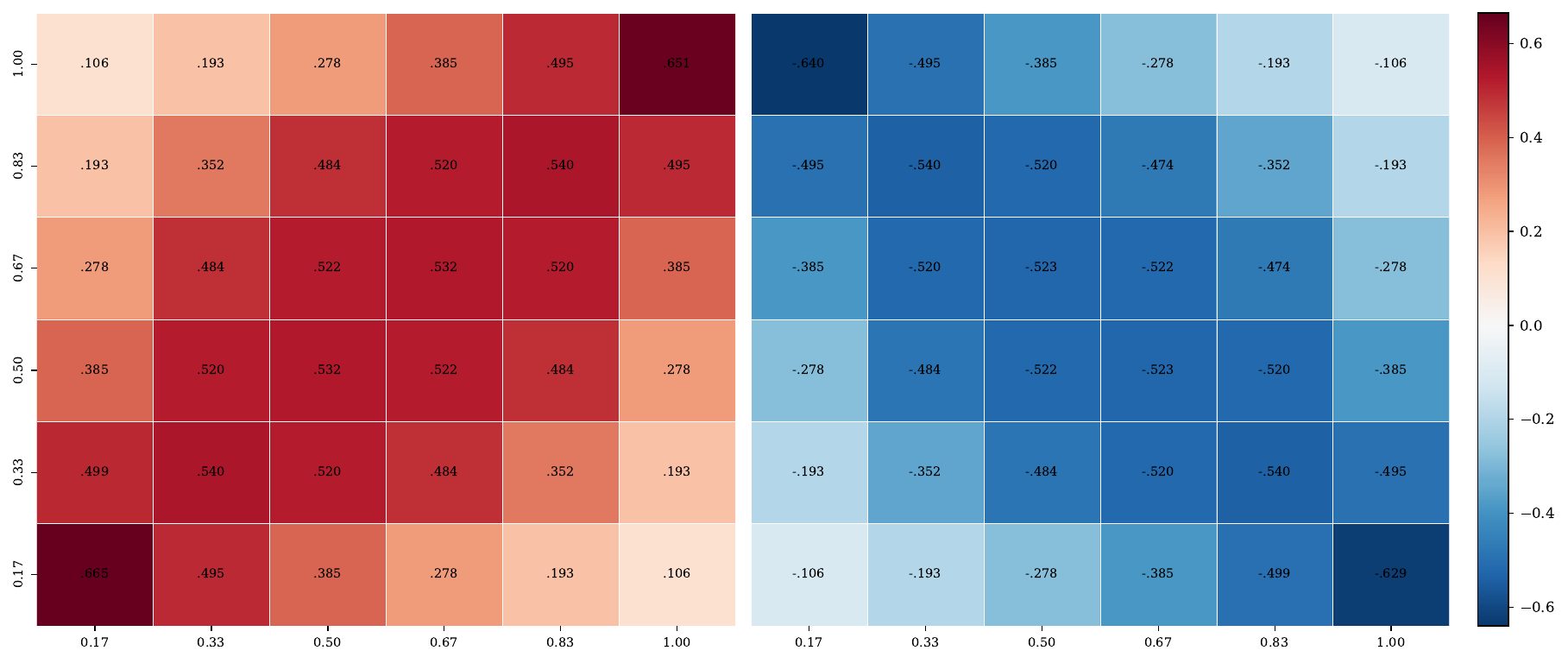}
           
            \end{subfigure}

            \begin{subfigure}{\linewidth}
                \centering
                \includegraphics[width=0.75\linewidth]{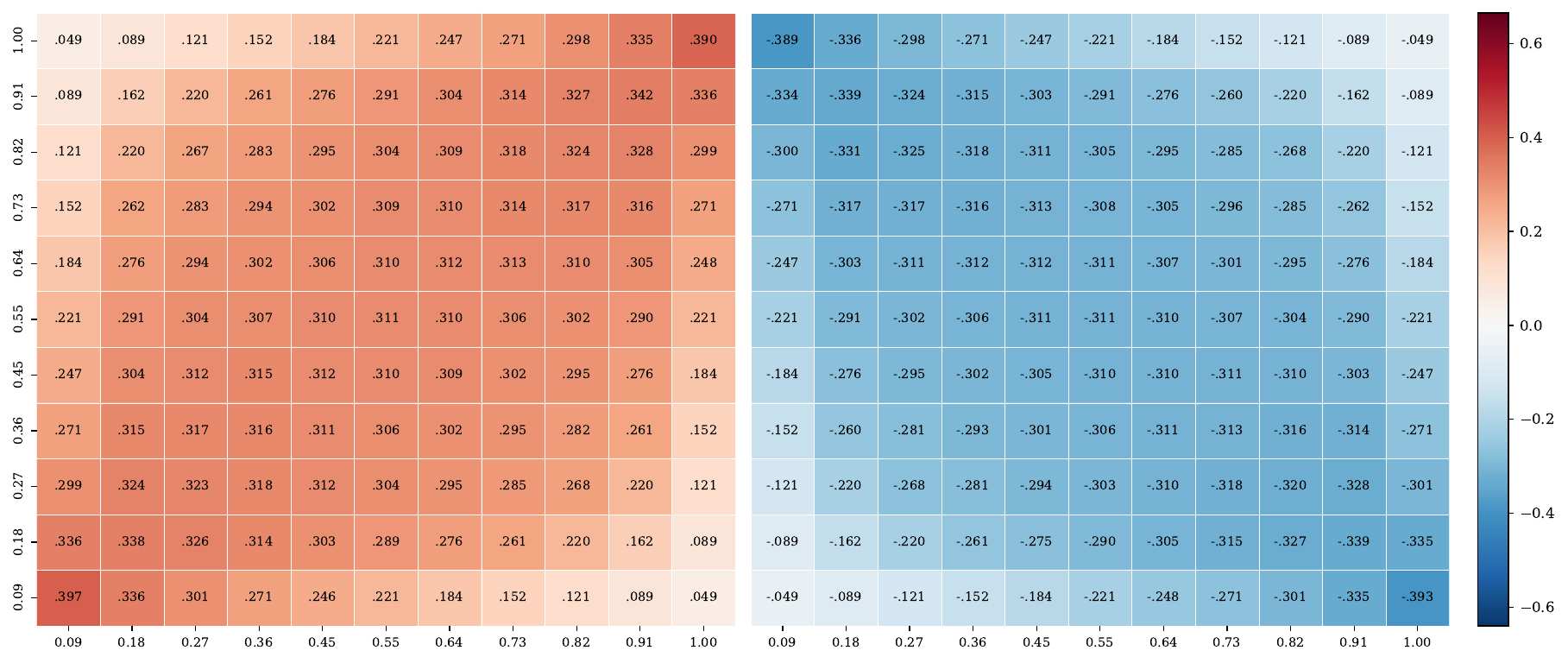}
           
            \end{subfigure}

            \begin{subfigure}{\linewidth}
                \centering
                \includegraphics[width=0.75\linewidth]{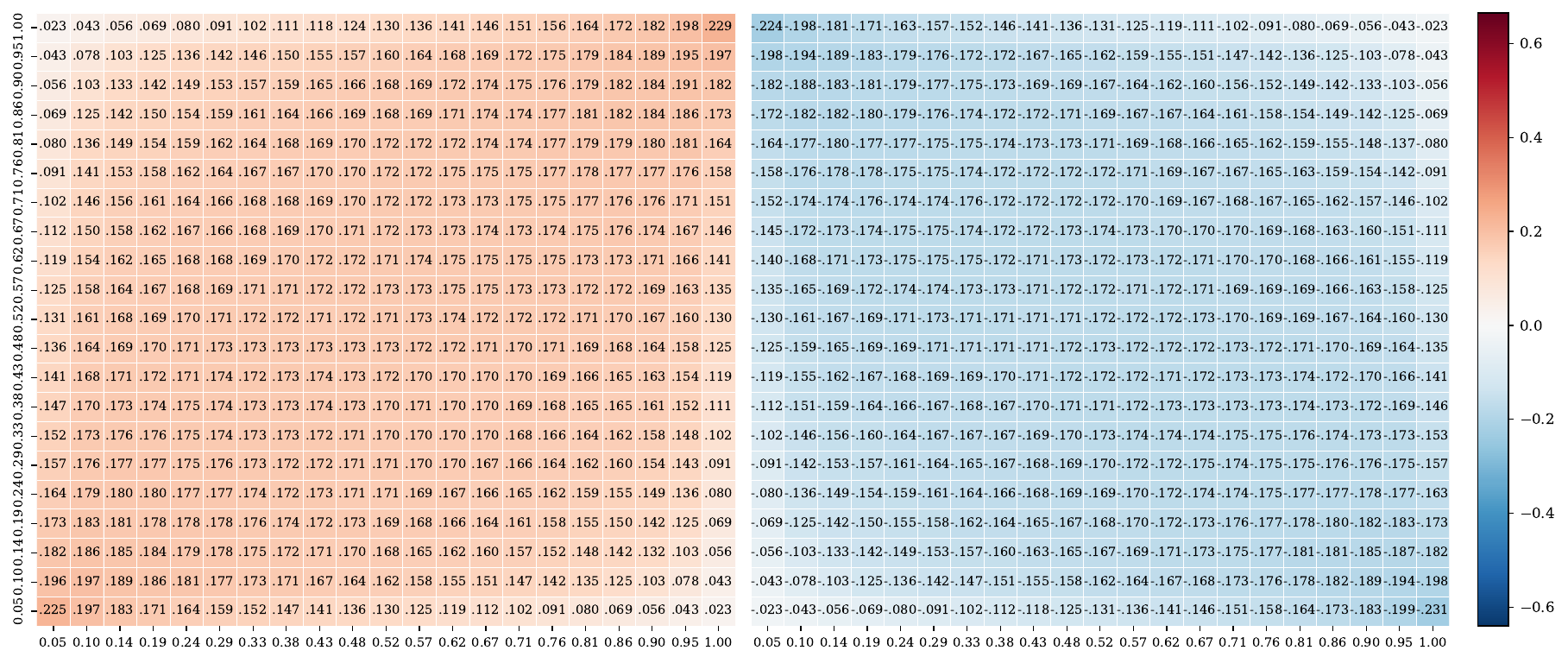}
               
            \end{subfigure}
        \end{minipage}
    }
    \caption{Critical surfaces (from the left): upper $U_{k,\eta_{k,n}(5\%)}$ and lower $L_{k,\eta_{k,n}(5\%)}$ which are calculated for different sample size under $H_0$ using $10^6$ Monte-Carlo iterations. From top to bottom: $n=25$ ($k=6$, $\eta_{6,25}(5\%) = 0.3613\%$), $n=100$ ($k=11$, $\eta_{11,100}(5\%)=0.0977\%$) and $n=400$ ($k=21$, $ \eta_{21,400}(5\%)=0.0378\%$).}
    \label{fig:u_n and l_n}
\end{figure}

We can now define the global independence test and the post-hoc local dependence test using $L_{\eta_n(\alpha)},$ $U_{\eta_n(\alpha)}$ from formula \eqref{kryt_lh}. The global test result is 
$$
G_{n}=1-\prod\limits_{i=1}^{n-1} \prod\limits_{j=1}^{n-1} 1\left(L_{\eta_n(\alpha)}(i/n,j/n)\leq q_n(i/n,j/n)\leq U_{\eta_n(\alpha)}(i/n,j/n)\right).
$$
The local positive dependence area is
$$D^+_n=\bigcup\limits_{(i,j): q_n(i/n,j/n)>U_{\eta_n(\alpha)}(i/n,j/n)} \left(\frac{i-0.5}{n},\frac{i+0.5}{n}\right]\times\left(\frac{j-0.5}{n},\frac{j+0.5}{n}\right],$$
and the local negative dependence area is
$$D^-_n=\bigcup\limits_{(i,j): q_n(i/n,j/n)<L_{\eta_n(\alpha)}(i/n,j/n)} \left(\frac{i-0.5}{n},\frac{i+0.5}{n}\right]\times\left(\frac{j-0.5}{n},\frac{j+0.5}{n}\right].$$
In Section \ref{dowod} we prove the following theorem.
\begin{tw}\label{twierdzenie}
 Let $(X_1,Y_1),\ldots,(X_n,Y_n)$ be i.i.d. from the distribution with continuous margins. Then the test $(G_{n},D^+_n,D^-_n)$ is consistent in the following way
 \begin{align}\label{TwP1}
 P_{H_0}(G_{n}=1 \ \ \vee \ \  D^+_n \neq \emptyset \ \  \vee \ \  D^-_n \neq \emptyset)\leq\alpha
 \end{align}
 \begin{align}\label{TwP2}
 \lim\limits_{n\rightarrow \infty} P_{H_1}(G_{n}=1)=1,
 \end{align}
\begin{align}\label{TwP3}
\forall {(u,v): q(u,v)>0} \ \lim\limits_{n\rightarrow \infty} P_{H_1}((u,v)\in D^+)=1, \ \ \ \forall {(u,v): q(u,v)<0}\  \lim\limits_{n\rightarrow \infty} P_{H_1}((u,v)\in D^-)=1. 
\end{align}
\end{tw}
Although the test $(G_{n},D^+_n,D^-_n)$ is consistent, one can construct the more practical version with appropriate smoothing. The distribution of $q_n(i/n,j/n)$ is very inconvenient for $(i/n,j/n)$ close to the edge of the unit square $[0,1]^2$. For example $q_n(1/n,1/n)$ can only take 2 values: $-1/(n-1)$ and $1$ with probabilities, under $H_0$, $(n-1)/n$ and $1/n$, respectively. As a consequence, for any $\alpha<1$ and $n$ large enough, we have $L_{\eta_n(\alpha)}=-1/(n-1)$ and $P_{H_0}(q_n(1/n,1/n) < L_{\eta_n(\alpha)}(1/n,1/n))=0$ so that we cannot observe a negative dependence in the vicinity of that point. Our solution to this problem is the following smoothed version of the test. 
We define a new grid 
\[
d_{k} = \left\{\left(\frac{s}{k}, \frac{t}{k}\right): s,t\in\{1,\ldots,k\}\right\},
\]
where $k < n$ is any positive integer number. Based on this partition, we obtain $k^2$ rectangles. We denote them by $M_{st} = ((s-1)/k,s/k] \times ((t-1)/k,t/k],$ as $s,t\in\{1,\ldots,k\}.$
The new version of the estimator of $q$, say $q_{k,n}$, is calculated as the mean of values of $q_n$ in every rectangle, i.e., for any $(u,v)\in(0,1)^2$
\begin{align}\nonumber
(s,t)=\left(\lceil uk\rceil,\lceil vk\rceil\right), \ \ \ (u_s,v_t)=\left(\frac{\lceil uk\rceil}{k},\frac{\lceil vk\rceil}{k}\right),\\
q_{k,n}(u,v)=q_{k,n}(u_s,v_t) = \frac{\sum\limits_{(i,j): (i/n,j/n)\in M_{st}} q_n(i/n,j/n)}{\sum\limits_{(i,j): (i/n,j/n)\in M_{st}} 1}. \label{formula_smooth}
\end{align}
For example, provided that $k = \lfloor\sqrt{n}\rfloor$ and $n=100,$ we obtain $100$ rectangles with $100$ values of $q_n$ in each. Empirical simulations show that choosing $k$ of the order of $\sqrt{n}$ yields satisfactory results in terms of power. In Section~5 we propose a division scheme for a less dense grid, in which the value of k is obtained as a by-product. Figure \ref{fig:histograms_bar_q} illustrates how this procedure, based on averaging, results in smoothing the point distributions of the estimator, particularly near the edges of the unit square. 
Let us introduce the appropriate quantiles $L_{k,\eta}$, $U_{k,\eta}$  
$$P_{H_0}\Big(q_{k,n}(u_s,v_t) < L_{k,\eta}(u_s,v_t)\Big)\leq\eta/2, \ \ P_{H_0}\Big(q_{k,n}(u_s,v_t) \leq L_{k,\eta}(u_s,v_t)\Big)\geq\eta/2,$$
$$P_{H_0}\Big(q_{k,n}(u_s,v_t)) > U_{k,\eta}(u_s,v_t)\Big)\leq\eta/2,\ \  P_{H_0}\Big(q_{k,n}(u_s,v_t) \geq U_{k,\eta}(u_s,v_t)\Big)\geq\eta/2,$$
and the local significance level:
\begin{align}
\eta_{k,n}(\alpha) = {} & \sup\Bigg\{\eta\in[0,\alpha]: \nonumber \\
& P_{H_0}\Bigg(\exists_{(s,t)\in\{1,\dots,k\}^2}:
\Big\{L_{k,\eta}\left(u_s,v_t\right)>q_{k,n}\left(u_s,v_t\right)\Big\}
\cup
\Big\{q_{k,n}\left(u_s,v_t\right)>U_{k,\eta}\left(u_s,v_t\right)\Big\}\Bigg)
\leq \alpha \Bigg\}. \nonumber
\end{align}
The global test for independence and the post-hoc test for local dependence using a smoothed version are as follows
$$
G_{k,n}=1-\prod\limits_{s=1}^{k} \prod\limits_{t=1}^{k} 1\left(L_{k,\eta_{k,n}(\alpha)}(u_s,v_t)\leq q_{k,n}(u_s,v_t) \leq U_{k,\eta_{k,n}(\alpha)}(u_s,v_t)\right).
$$
The local positive and local negative dependence areas are
$$D^+_{k,n}=\bigcup\limits_{(s,t): q_{k,n}(u_s,v_t)>U_{k,\eta_{k,n}(\alpha)}(u_s,v_t)} M_{st},\ \ \ \ \ \ D^-_{k,n}=\bigcup\limits_{(s,t): q_{k,n}(u_s,v_t)<L_{k,\eta_{k,n}(\alpha)}(u_s,v_t)} M_{st}.$$
The surfaces $U_{k,\eta_{k,n}(\alpha)}$ and $L_{k,\eta_{k,n}(\alpha)}$, calculated using the Monte-Carlo method, are presented in Figure \ref{fig:u_n and l_n}. 
\\
\\

\noindent One can see that the surfaces have smaller values for larger sample sizes. In the proof of Theorem \ref{twierdzenie} we will show that critical surfaces tend to zero in every fixed point $(u,v)\in(0,1)^2$. So, for any alternative $H_1$, the probability of exceeding one of the two critical surfaces, in a point where $q(u,v)\neq 0$, tends to 1 due to the consistency of the estimator.

\begin{wniosek}\label{wniosek}
 Let $(X_1,Y_1),\ldots,(X_n,Y_n)$ be i.i.d. from the distribution with continuous margins. Then the test $(G_{k,n},D^+_{k,n},D^-_{k,n})$ is consistent in the same way as in Theorem \ref{twierdzenie} for any $k=k(n)$ such that $1<k(n)\leq n$ and $\lim\limits_{n \rightarrow \infty} k(n)=\infty.$
\end{wniosek}
\begin{obs}
Note that the weight in the denominator of function $q$ is very important in the smoothed version of the test. The numerator of $q$ has a different variance at each point $(u,v)$ and this should be taken into account during averaging. The denominator is a variance-stabilizing weight which makes the averaging more fair. 
\end{obs}

\begin{obs}
In theorems concerning the consistency of copula-based independence tests, additional regularity conditions on the partial derivatives of the true copula are usually assumed. These conditions are required for results on the weak convergence of the empirical process. In our approach, however, we do not rely on such theorems and therefore we do not need any additional regularity assumptions. This makes our test more general.  
\end{obs}

\newpage
\section{Examples and numerical results}
\subsection{COVID dataset}
\hspace*{\parindent} The first empirical illustration of our methodology utilizes data from the COVID‑19 pandemic caused by the SARS‑CoV‑2 virus. The dataset comprises country‑level statistics, including daily confirmed infection counts, as well as macroeconomic indicators such as population size and Gross Domestic Product (GDP). Details on the dataset construction and data sources are provided in \cite{HasellJoe2020Acdo}. 

To ensure comparability with prior work, we employ the same preprocessed version of the dataset as in \cite{CmielAdamM.2021Asmt}. In line with their procedure, we restrict the analysis to the 73 countries for which all relevant variables are fully observed, excluding the remaining 7 countries due to missing data.
\begin{figure}[H]
\centering
    \includegraphics[width=0.9\linewidth]{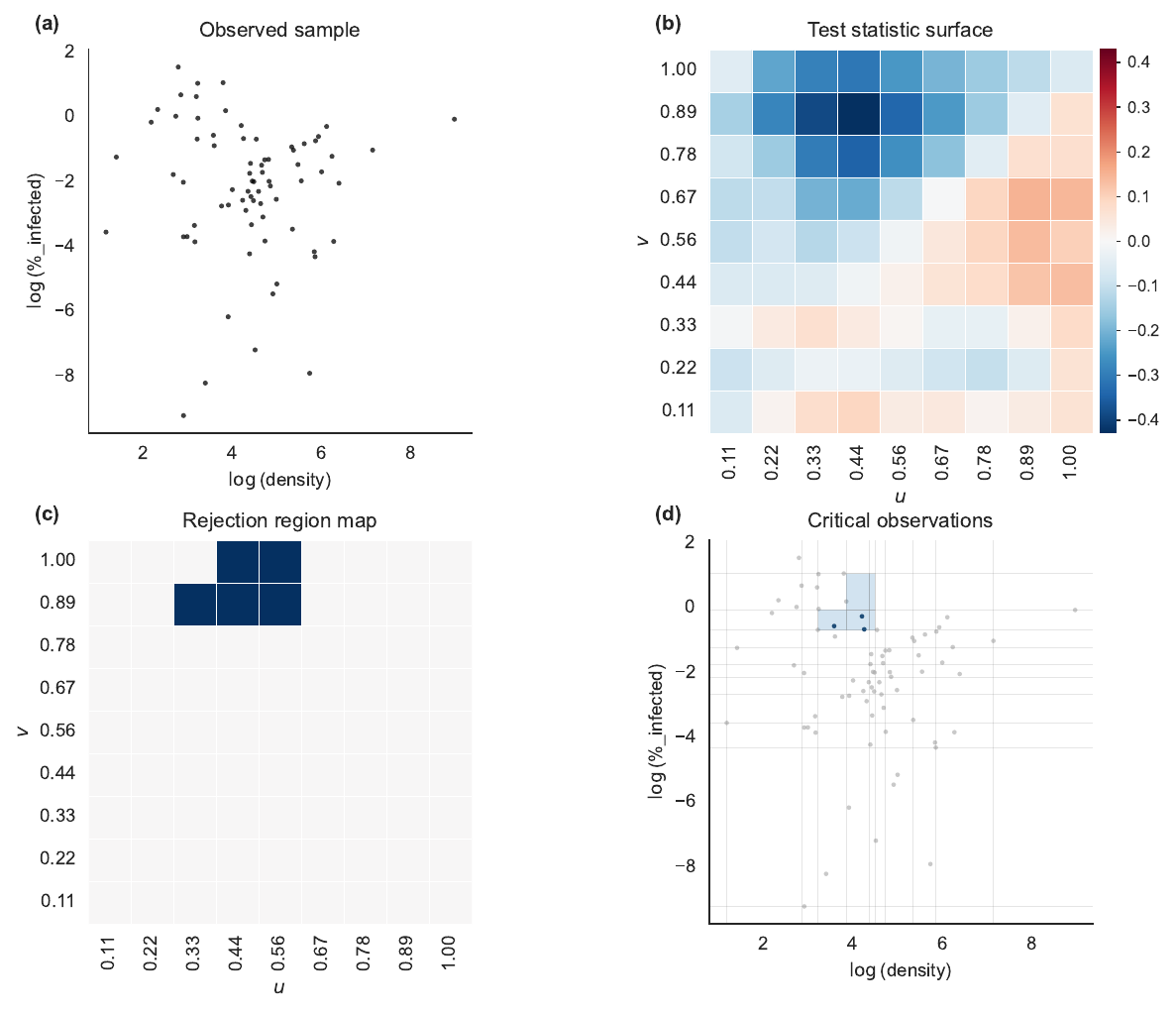}
    \caption{Panels from the upper left show: (a) the observed data, with population density plotted against the percentage of infected individuals after logarithmic transformation; (b) a heatmap of the smoothed estimator $q_{k,n}$; (c) a heatmap indicating the regions in which the critical surfaces are significantly exceeded; and (d) the observed data with the points identified in the rejection map highlighted. The calculations come from $10^6$ Monte Carlo simulations with $k=9$ and $\eta_{9,73}(5\%) \cong 0.1367\%.$}
    \label{fig:COVID_dataset}
\end{figure}
They examine the relationship between national population density and the proportion of infected individuals by employing an independence test derived from the quantile dependence function introduced in \cite{Ledwina2015}. Furthermore, \cite{cmiel2024detectingdependencestructurevisualization} demonstrate that both the dCov and Max BET tests fail to detect this relationship, as they do not reject the null hypothesis of independence. Our local test successfully identifies the underlying dependence structure and rejects the hypothesis of independence, as depicted in Figure \ref{fig:COVID_dataset}.

\subsection{Danish fire insurance dataset}
\hspace*{\parindent} We consider the Danish fire insurance dataset originally presented in \cite{McNeilAlexanderJ.1997EtTo}. A more detailed description and further analyses are also provided in \cite{EmbrechtsPaul1997MEEF}. The data are available in the R package \texttt{fitdistrplus} and comprise 2167 fire loss observations recorded over the period 1980–1990. 

\begin{figure}[H]
\centering
    \includegraphics[width=0.9\linewidth]{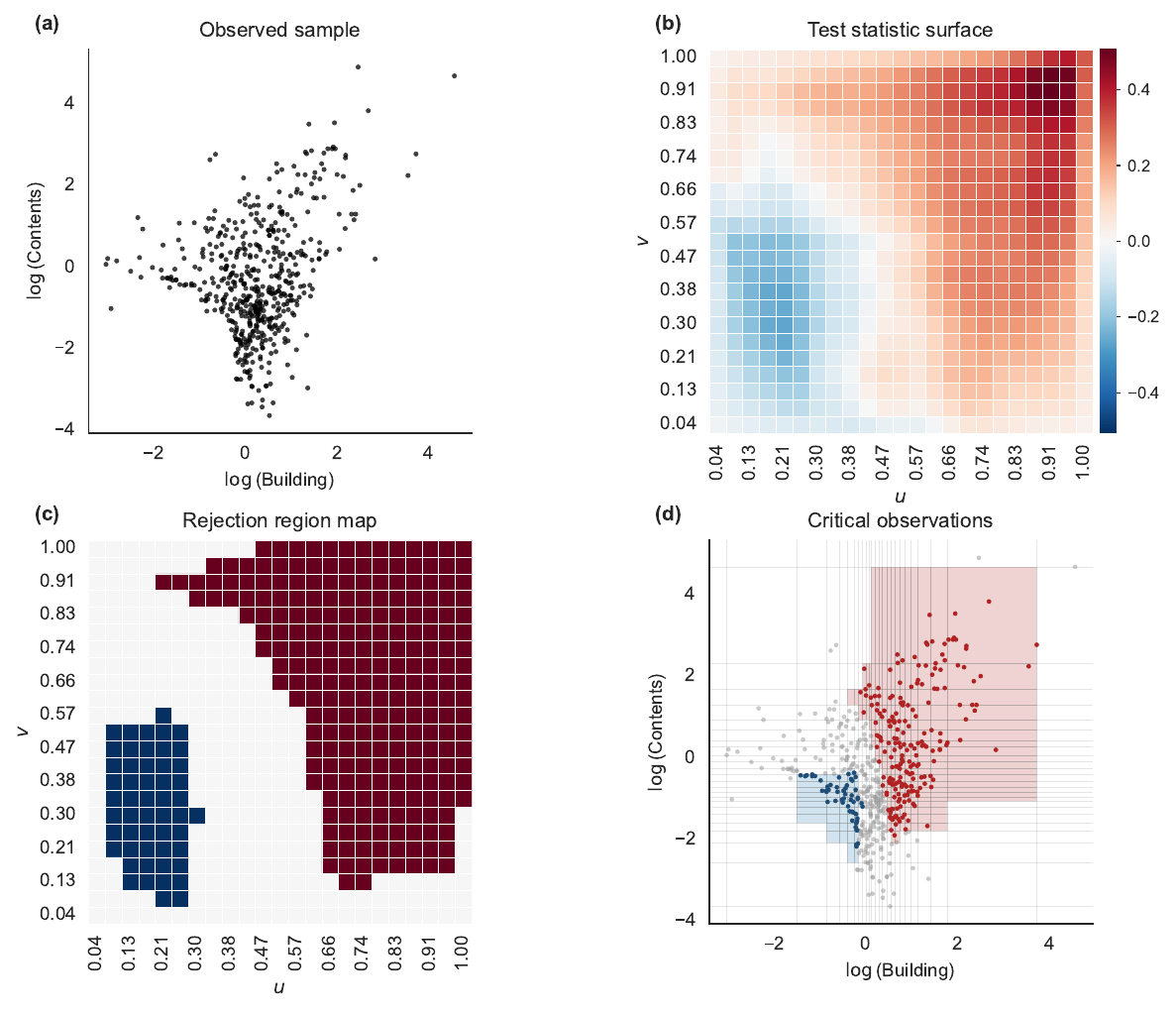}
    \caption{Panels from the upper left show: (a) observed data, with log-transformed content loss plotted against log-transformed building loss; (b) a heatmap of the smoothed estimator $q_{k,n}$; (c) a heatmap showing regions where the critical surfaces are significantly exceeded; and (d) the observed data with points highlighted as identified in the rejection map. Results are based on $10^6$ Monte Carlo simulations with $k=23$ and $\eta_{23,517}(5\%) \cong 0.03296\%.$}
    \label{fig:danish_ins}
\end{figure}

In the present study, we focus on two variables: \emph{Buildings} and \emph{Contents}, which correspond to the total loss amounts under the building and contents coverages, respectively. To reduce the sample size and concentrate on strictly positive claims, we restrict attention to claims with non-zero amounts in buildings, contents, and profit losses. This filtering yields a subsample of 517 observations, depicted in Figure \ref{fig:danish_ins}. 

The same dataset was examined in \cite{cmiel2024detectingdependencestructurevisualization}, where the authors assessed the null hypothesis of independence between the considered variables using the global test statistic $T_n$.   
Based on their results, it is clear that the null hypothesis can be rejected. Using their approach, we can also determine where the local dependence is significant, but without controlling the overall significance level. Our test addresses this problem. As shown in the third and fourth plots in Figure \ref{fig:danish_ins}, the real type of this relationship can be seen. It is clearly visible that there is a local negative dependence between contents losses and small building losses, which may indicate a situation where some contents losses do not lead to severe building destructions. However, there is a positive relationship between building losses that exceed a certain loss amount, which can mean that extensive fire damages that cause huge destruction of the building are related to greater contents damages.    

\subsection{Comparison of Empirical Powers}
\hspace*{\parindent}We now compare the independence test proposed in this paper with the one introduced in \cite{cmiel2024detectingdependencestructurevisualization}, which is also based on the quantile dependence function. For comparability, we adopt exactly the same data-generating distributions as in \cite{cmiel2024detectingdependencestructurevisualization}. In our experiments, we set the sample size to $n=100$ and $k=11$. The model notation is retained from the referenced study without modification.

The following Table \ref{tab:power} reports the empirical powers for the two tests under consideration, namely $T_n$ (as defined in \cite{cmiel2024detectingdependencestructurevisualization}) and $G_{k,n}$ (proposed in the present work), estimated from $10^4$ Monte Carlo replications. For additional benchmarking, we also include the BET test of \cite{ZhangKai2019BoI}, implemented via its Max BET variant with $d_{\max}=4.$ Within this testing framework, the realized (effective) significance level typically falls below the nominal \(\alpha\) threshold as a consequence of applying multiple-testing adjustments. To enable a fair comparison across procedures and to increase the statistical power while preserving the prescribed type I error rate, we perform a calibration of the test under the null hypothesis. Concretely, we approximate the distribution of p-values under the assumption of independence and derive the critical cutoff that corresponds to the target significance level. For benchmarking purposes, we additionally include the distance covariance (dCov) test proposed by \cite{SzekelyGaborJ.2009BDC} in our comparison.
\begin{table}[H]
\centering
\resizebox{0.9\textwidth}{!}{%
\begin{tabular}{
  |>{\raggedright\arraybackslash}m{0.4\textwidth}|
   >{\centering\arraybackslash}m{0.12\textwidth}|
   >{\centering\arraybackslash}m{0.12\textwidth}|
   >{\centering\arraybackslash}m{0.12\textwidth}|
   >{\centering\arraybackslash}m{0.12\textwidth}|
}
\hline
\textbf{Model} & \textbf{dCov}  & \textbf{$G_{k,n}$-test} & \textbf{$T_n$-test} & \textbf{Max BET} \\
\hline
\rowcolor{lightgreen}
\multicolumn{5}{|l|}{\textbf{Simple Regression}} \\
\hline
\hspace{1em}SR1: Linear & 0.750 & 0.582 & 0.668 & 0.318\\
\hspace{1em}SR2: Root & 0.794 & 0.658 & 0.753 & 0.333\\
\hspace{1em}SR3: Step & 0.841 & 0.692 & 0.765 & 0.531\\
\hspace{1em}SR4: Logarithmic & 0.258 & 0.298 & 0.395 & 0.222\\
\hspace{1em}SR5: W-shaped & 0.382 & 0.355 & 0.407 & 0.416 \\
\hline
\rowcolor{lightgreen}
\multicolumn{5}{|l|}{\textbf{Heteroscedastic Regression}} \\
\hline
\hspace{1em}HR1: Reciprocal & 0.112 & 0.616 & 0.676 & 0.092\\
\hspace{1em}HR2: Linear & 0.342 & 0.453 & 0.465 & 0.284 \\
\hline
\rowcolor{lightgreen}
\multicolumn{5}{|l|}{\textbf{Random-Effect-Type Models}} \\
\hline
\hspace{1em}RE1: Linear & 0.605 & 0.671	 & 0.703 & 0.256 \\
\hspace{1em}RE2: Quadratic & 0.339 & 0.794 & 0.946 & 0.460\\
\hspace{1em}RE3: Reciprocal & 0.109 & 0.284	& 0.094 & 0.642\\
\hspace{1em}RE4: Heavy tailed & 0.106 & 0.551 & 0.803 & 0.175 \\
\hline
\rowcolor{lightgreen}
\multicolumn{5}{|l|}{\textbf{Classical Bivariate Models}} \\
\hline
\hspace{1em}BM1: Gaussian & 0.774 & 0.624 & 0.731 & 0.334 \\
\hspace{1em}BM2: Mixture I & 0.230 & 0.601 & 0.643 & 0.090\\
\hspace{1em}BM3: Mixture II & 0.075 & 0.466	& 0.608 & 0.080\\
\hspace{1em}BM4: Switched regression & 0.192 & 0.657 & 0.690 & 0.088\\
\hspace{1em}BM5: Mai-Scherer copula & 0.570 & 0.697	& 0.693 & 0.164\\
\hspace{1em}BM6: Gumbel copula & 0.548 & 0.447	& 0.486 & 0.194\\
\hspace{1em}BM7: Gumbel-Hougaard copula & 0.651 & 0.559	& 0.675 & 0.234\\
\hspace{1em}BM8: Cauchy & 0.209 & 0.817 & 0.988 & 0.700\\
\hspace{1em}BM9: Student symmetric & 0.089 & 0.346	& 0.549 & 0.159\\
\hspace{1em}BM10: Student skew & 0.261 & 0.580 & 0.678 & 0.149\\
\hspace{1em}BM11: Sub-Gaussian & 0.173 & 0.388	& 0.551 & 0.102\\
\hline

\end{tabular}
}	

\caption{The comparison of empirical powers between tests dCov, $G_{k,n},T_n$ and Max BET under different alternatives.}
\label{tab:power}
\end{table}
Comparing the results presented in Table \ref{tab:power} it is clearly visible that our test based on critical surfaces has slightly lower empirical powers than the global test $T_n$ in almost every case. It is consistent with our expectations. In general, post-hoc tests are less powerful than their global analogues due to their locality. It should be noted that our test shows better performance for data from the RE3 model. The $T_n$-test is more powerful than the other similar tests, particularly when the dependence structure is close to the edge of the unit square. However, in this case, the dependence is located near the centre of the square side. This particular place is a weakness of the $T_n$-test which has trouble capturing the dependencies in this area. Our test detects this type of dependence more frequently because it treats every area of the partition equally. 

The Max BET-test exhibits the lowest power in almost all cases, mainly due to the use of the Bonferroni correction for multiple testing problem. In this method, the test examines all possible binary resolutions up to the $d_{max}$ level of refinement. Across these resolutions, it evaluates the corresponding bit-level interactions and identifies the one that shows the strongest discrepancy from independence. Since the procedure searches over multiple candidate interactions, the resulting p-value is adjusted for multiple testing, using the Bonferroni correction. For a large depth $d$ this multi testing adjustment makes rejecting the null hypothesis of independence increasingly difficult. 

Nevertheless, if the type of expected dependence is known in advance, an appropriate choice of the depth parameter can increase the power of the test and make it more sensitive. This explains why the power is particularly high in the RE3 case. In contrast, our proposed test performs well even when no assumptions about the form of dependence are made, as it is able to detect such dependence automatically.

It should also be noted that the critical values of the $T_n/\sqrt{n}$ statistic are very close to the average values of the critical surfaces. It is shown in the Table \ref{tab:comp_table}. The division of ${T_n}$ is due to the definition of this statistic. Both surfaces tend to zero under the independence and are based on the quantile dependence function; therefore, this behaviour is expected.

\begin{table}[h]
\centering
\resizebox{.8\textwidth}{!}{%
\begin{tabular}{|c|c|c|c|c|c|}
\hline
\textbf{Sample size} & $\bm{T_n/\sqrt{n}}$ & $\bm{\max U_{k,\eta_{k,n}}}$ & $\bm{\min L_{k,\eta_{k,n}}}$ & $\bm{\mathrm{mean} \ U_{k,\eta_{k,n}}}$ & $\bm{\mathrm{mean} \ L_{k,\eta_{k,n}}}$ \\
\hline
\hline
$n=25$ & 0.5290 & 0.6316 & -0.6419 & 0.4079 & -0.4112\\
\hline
$n=100$ & 0.2800 & 0.3975 & -0.3932 & 0.2682 & -0.2681\\
\hline
$n=400$ & 0.1414 & 0.2292 & -0.2311 & 0.1590 & -0.1589\\
\hline
$n=500$ & 0.1268 & 0.2007 & -0.2021 & 0.1438 & -0.1436 \\
\hline
\end{tabular}}
\caption{Comparison of the critical values of the statistics \(T_n/\sqrt{n}\) with the corresponding critical surfaces for selected sample sizes (computed under a 5\% significance level). }
\label{tab:comp_table}
\end{table}

\section{The algorithm}
\hspace*{\parindent} In this section, we introduce an algorithm for the construction of critical surfaces and employ it to assess global independence and local dependence. Let \((X_1,Y_1),\ldots,(X_n,Y_n)\) denote an observed bivariate sample of size \(n\). Since it is sufficient to evaluate the estimator on a discrete set of points, we restrict our computations to the grid generated by \( \{1/n, 2/n, \ldots, (n-1)/n\}\), as previously described. The testing procedure can be summarized in the following steps:
\begin{itemize}
    \item Using the observed sample, compute the \textit{plug-in} estimator of the quantile dependence function, denoted by $q_n$, on the grid $\{1/n, 2/n, \ldots, (n-1)/n\}\times \{1/n, 2/n, \ldots, (n-1)/n\}$.
    \item For a fixed integer $k$, build a coarser grid $\Pi_k = \big\{M_{st} : s,t \in \{1,\ldots,k\}\big\}$ and compute the smoothed estimator values $q_{k,n}$ as the arithmetic mean of the points $q_n$ that fall within each cell $M_{st}$ of $\Pi_k$, according to the formula \eqref{formula_smooth} in Section 3.
    
    \item Due to the unknown distribution of the estimator \(q_{k,n}\), it is necessary to approximate its sampling distribution using Monte Carlo simulation for each region in \(\Pi_k\), and then compute the corresponding critical values at each point to compare them with the observed value of the statistic. Since the procedure is distribution-free, generate \(MC\) independent bivariate samples of size \(n\) from an uniform distribution. For each sample, evaluate the estimator \(q_{k,n}\), thereby obtaining Monte Carlo replications \(q_{k,n}^{*1},\ldots,q_{k,n}^{*MC}\).

    \item The next step is to estimate the local significance level. We initialize the auxiliary parameter \(\eta^{(0)} = \alpha/2\). In the $m$-th step, for each point $(u_s,v_t)$, we compute the empirical quantiles \(\eta^{(m)}/2\) and \(1-\eta^{(m)}/2\) of the Monte Carlo replications to obtain the lower and upper critical surfaces \(L_{k,\eta^{(m)}}\) and \(U_{k,\eta^{(m)}}\), respectively. The value of \(\eta^{(m)}\) is then iteratively adjusted, for example by the bisection method, until the resulting global significance level agrees with the target level $\alpha$ up to a prescribed tolerance $\varepsilon$. The final value is denoted by $\eta_{k,n}(\alpha).$
\end{itemize}

It is important to note that a grid $\Pi_k$ composed of regions each containing an identical number of points rarely exists. Consequently, certain cells may contain a higher or lower number of points than others. To alleviate this imbalance, we propose a more equitably distributed partition. For $k=\lfloor\sqrt{n-1}\rfloor$, define $r = \frac{k}{n-1}$ and $N = \left\lfloor\frac{n-1-k}{2k}\right\rfloor$. The division points are then given by:
\[
\Pi_k = \{ir, 1-ir: \ i = 1,\ldots,N\} \cup B, \quad \text{where } 
B =
\begin{cases}
\{0.5\} & \text{when } 0.5-Nr < r, \\
\emptyset & \text{otherwise.}
\end{cases}
\]
Under the adopted grid discretization, whenever an exactly uniform partition is not feasible, the remaining points are assigned to one of the central intervals (i.e. to one or two middle columns of the grid). The boundary cells thus retain identical numbers of subdivision points. Consequently, the overall allocation remains balanced, since each pair of corresponding cells is evaluated using averages computed from an equal number of points.

\begin{figure}[H]
    \centering
    \includegraphics[width=0.8\linewidth]{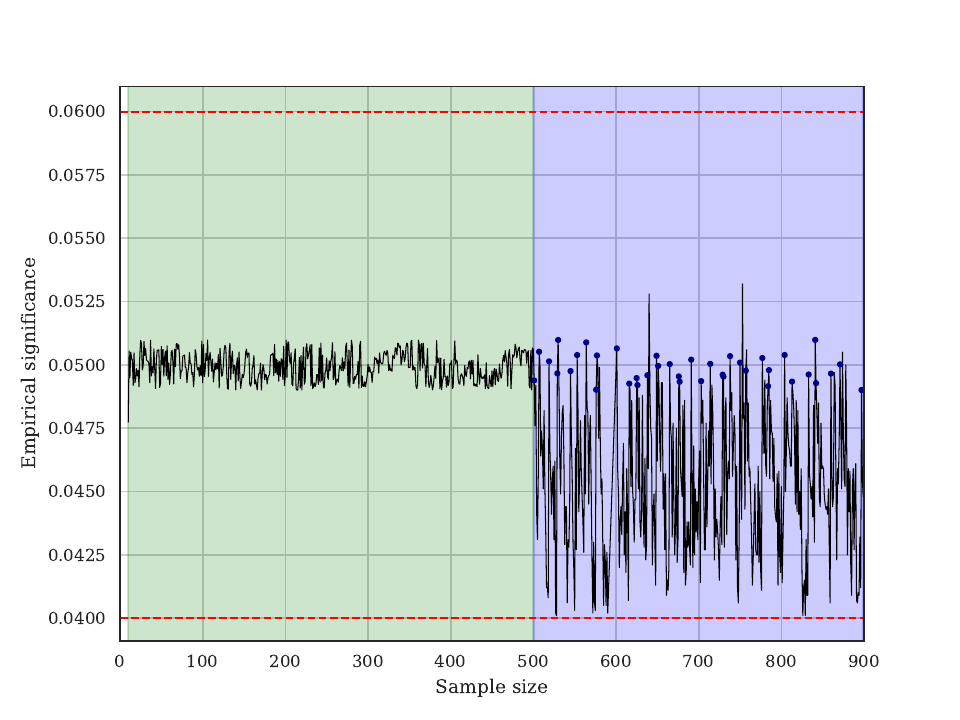}
    \caption{For sample sizes $n=10,\ldots,500$ the critical surfaces are computed separately for each value of $n$ (light green area). For larger sample sizes, however, they are computed only at larger intervals, as indicated by the blue points in the plot, under the constraint that the empirical significance level (using the closest surfaces) must remain within the range $\alpha\pm 1\%$ (purple area). If the lower boundary is exceeded, the new surfaces are determined.}
    \label{fig:emp_sign_alg}
\end{figure}

There is a further rationale for introducing this modification. Due to the computational complexity of the procedure, the determination of critical surfaces becomes increasingly time-consuming. For a sample of size $n$, it is necessary to compute $MC$ matrices of dimension $(n-1)\times(n-1)$, then obtain a smoothed version of the estimator by averaging and finally determine the local significance level. As $n$ increases, this sequence of operations becomes progressively more computationally demanding. The proposed partition scheme enables the computation of critical surfaces for selected values of $n$, which can then be applied to nearby sample sizes. This obviates the need to determine a separate critical surfaces for each individual $n$, albeit at the cost of a reduction in statistical power. This effect is illustrated in Figure \ref{fig:emp_sign_alg}. Once the sample size exceeds $500$, a decline in empirical power can be observed, with only isolated peaks occurring at the sample sizes for which the critical surfaces have been explicitly computed.

Another aspect of practical implementation that deserves particular attention is the numerical accuracy of the computations and the choice of the number of Monte Carlo replications. Because the local significance levels $\eta_{k,n}(\alpha)$ are very small, a sufficiently large number of samples from the empirical distribution is required to obtain reliable estimates of the corresponding quantiles (in our simulations, we set $MC = 10^6$). Inadequate numerical precision or an insufficient number of replications can result in unstable estimates of the critical values. Consequently, to ensure maximal numerical accuracy and reproducibility of the results, we recommend carrying out all computations in double-precision arithmetic, i.e., using the float64 format.
The Python code used to generate and analyze the simulation results presented in this study is publicly available at: \url{https://github.com/PostHocDepAuthors/The-post-hoc-test-for-local-dependence}. The repository contains the source code, input/configuration files, and a README file describing how to install the package.

\newpage
 
\newpage
\section{Proof of Theorem \ref{twierdzenie} and Corollary \ref{wniosek}}\label{dowod}
The construction of critical surfaces (\ref{alpha}) guarantees that (\ref{TwP1}) is true. To show (\ref{TwP2}) and (\ref{TwP3}) it is enough to prove that 
$$\forall (u,v)\in(0,1)^2  \ \ \lim\limits_{n \rightarrow \infty} L_{\eta_n(\alpha)}(u,v)= \lim\limits_{n \rightarrow \infty} U_{\eta_n(\alpha)}(u,v)=0,$$
since for any fixed $(u,v)$ the following implications are true
$$\left[q(u,v)>0, \ \ \ q_n(u,v)\rightarrow^{P}q(u,v), \ \ \ 0\leq U_{\eta_n(\alpha)}(u,v)\rightarrow 0\right]\Rightarrow P(q_n(u,v)>U_{\eta_n(\alpha)}(u,v))\rightarrow 1,$$
$$\left[q(u,v)<0, \ \ \ q_n(u,v)\rightarrow^{P}q(u,v), \ \ \ 0\geq L_{\eta_n(\alpha)}(u,v)\rightarrow 0\right]\Rightarrow P(q_n(u,v)<L_{\eta_n(\alpha)}(u,v))\rightarrow 1.$$
For any fixed $(u,v)\in(0,1)^2$ the lower critical surface $L_{\eta_n(\alpha)}(u,v)$ is a quantile of the order $\eta_n(\alpha)/2$ in the distribution of $q_n(u,v)=(C_n(u,v)-uv)/\sqrt{uv(1-u)(1-v)}$ under $H_0$. The exact distribution can be obtained from the following fact.
\begin{fakt}
For any fixed $(u,v)\in(0,1)^2,$ provided $H_0$ is true, holds 
\begin{align}
   P\big(nC_n(u,v) = k\big) = \frac{\bigbinom{\lfloor nv \rfloor}{k}\bigbinom{n-\lfloor nv \rfloor}{\lfloor nu \rfloor-k}}{\bigbinom{n}{\lfloor nu \rfloor}}, \label{rozklad_nCn}
   \end{align}
   if $k\in \big\{\max\{0,\lfloor nu \rfloor+\lfloor nv \rfloor-n\}, \ldots, \min\{\lfloor nu \rfloor,\lfloor nv \rfloor\}\big\},$ otherwise $P\big(nC_n(u,v) = k\big) = 0.$ It can be written briefly as $nC_n(u,v) \stackrel{H_0}{\sim} \mathrm{Hyp}(\lfloor nu \rfloor,\lfloor nv \rfloor, n).$
\end{fakt}
To evaluate the asymptotic behavior of $L_{\eta_n(\alpha)}(u,v)$ we will use the following proposition, which is proved in the end.
\begin{prop}\label{prop2}
  For every $(u,v)\in(0,1)^2$, $x\in\mathbb{R}, \ d>0$ we have 
  \[
  \left|P_{H_0}(C_n(u,v)-uv \leq  x) -\Phi\left(\sqrt{\frac{n}{d}}x\right)\right| \leq A_1\left(\sqrt{n}x^2+\frac{1}{\sqrt{n}}\right)\exp(-A_2 nx^2),
  \]
  where $A_1,A_2$ are some positive constants, independent of $x$ and $n$ (dependent of $(u,v)$).
\end{prop}
Using the above proposition, we can write
    \begin{align}
    \Bigg| P_{H_0}(q_n(u,v)\leq L_{\eta_n(\alpha)}(u,v))&-\Phi\Bigg(\sqrt{\frac{n}{d}}L_{\eta_n(\alpha)}(u,v)\Bigg)\Bigg| \nonumber\\
    &\leq C_1\left(\sqrt{n}L^2_{\eta_n(\alpha)}(u,v)+\frac{1}{\sqrt{n}}\right)\exp\left(- C_2nL^2_{\eta_n(\alpha)}(u,v)\right), \label{eq:th_1_begining_exp}
    \end{align}
    for some positive constants $C_1$ and $C_2.$ Suppose, for the sake of contradiction, that $L_{\eta_n(\alpha)}(u,v) < -1/\sqrt[4]{n} < 0$. We will show that it cannot be true.
    Using Proposition \ref{prop1} and the assumption $L_{\eta_n(\alpha)}(u,v) < -1/\sqrt[4]{n}$ we can bound the left side of the above inequality from below as follows
    \begin{align}
        \frac{K}{n^2}\leq \frac{\alpha}{2n^2} - \frac{\exp\left(-\sqrt{n}/(2d)\right)}{n^{1/4}} \leq \left| P_{H_0}(q_n(u,v)\leq L_{\eta_n(\alpha)}(u,v))-\Phi\left(\sqrt{\frac{n}{d}}L_{\eta_n(\alpha)}(u,v)\right)\right|, \label{eq:first_ineq}
    \end{align}
    for some positive constant $K\in(0,1).$ The first relation holds for sufficiently large $n$ and the second is due to the rapid (exponential) decay of the Gaussian tails. The right side of $\eqref{eq:th_1_begining_exp}$ can be upper-bounded as 
    \begin{align}
        C_1\left(\sqrt{n}L^2_{\eta_n(\alpha)}(u,v)+\frac{1}{\sqrt{n}}\right)\exp\left(- C_2nL^2_{\eta_n(\alpha)}(u,v)\right) \leq C_1\sqrt{n}\exp\left(-C_2\sqrt{n}\right) \leq C_1\exp\left(-C_2\sqrt{n}/2\right),
        \label{eq:second_ineq}
    \end{align}
    where the last inequality holds for sufficiently large $n$. Additionally, the first inequality follows from the fact that the support of $q_n(u,v)$  is contained in the interval $[-1,1].$ 
    
    In summary, using inequalities \eqref{eq:first_ineq} and \eqref{eq:second_ineq}, we obtain the following

    \[
    \frac{K}{n^2}\leq \left| P_{H_0}(q_n(u,v)\leq L_{\eta_n(\alpha)}(u,v))-\Phi\left(\sqrt{\frac{n}{d}}L_{\eta_n(\alpha)}(u,v)\right)\right|\leq C_1\exp\left(-C_2\sqrt{n}/2\right),
    \]
    and a contradiction arises because for large $n$, the right-hand side becomes strictly smaller than the left-hand side. Therefore,
    $$-1/\sqrt[4]{n} \leq L_{\eta_n(\alpha)}(u,v) <0,$$
   for large $n$ so $L_{\eta_n(\alpha)}(u,v)$ converges to zero as $n\rightarrow\infty$. In the same way, one can show that $U_{\eta_n(\alpha)}(u,v)$ converges to zero as $n\rightarrow\infty$ for any fixed $(u,v)\in(0,1)^2$.
\begin{obs} 
The lower bound $-1/\sqrt[4]{n} < L_{\eta_n(\alpha)}(u,v)$ in not optimal but good enough to prove consistency. Analysing the proof one can observe that for all $0<s<0.5$ and for all $(u,v)\in(0,1)^2$ there exists $n_0$ such that for all $n>n_0$  
$$-1/{n}^s \leq L_{\eta_n(\alpha)}(u,v) <0,\ \ \ \ 0< U_{\eta_n(\alpha)}(u,v) \leq1/{n}^s.$$
It is also obvious that $L_{\eta_n(\alpha)}(u,v)$ and $U_{\eta_n(\alpha)}(u,v)$ cannot tend to zero faster than $1/\sqrt{n}$ due to the weak convergence of $\sqrt{n}q_n(u,v)$ to $ N(0,1)$ under $H_0$ for any fixed $(u,v)\in (0,1)^2$.
\label{rem:rate_pow}
\end{obs}

   To prove Corollary \ref{wniosek} it is enough to observe that $\eta_{k,n}(\alpha)\geq\alpha/k^2\geq \alpha/n^2$ (smaller number of local tests). In addition, $q_{k,n}(u,v)$ is also asymptotically normal under $H_0$ with zero mean and even lower variance than $q_n(u,v)$ due to averaging. Using the fact that the quantile of the order $\alpha/n^2$ for $N(0, d/n)$ distribution tends to zero, one can show that $U_{k,\eta_{k,n}(\alpha)}(u,v)$ and $L_{k,\eta_{k,n}(\alpha)}(u,v)$ tend to zero for any fixed $(u,v)$ not slower than $U_{\eta_n(\alpha)}(u,v)$ and $L_{\eta_n(\alpha)}(u,v)$.

\subsection{Proof of the Proposition \ref{prop2}}
From the triangle inequality, we write     
\begin{flalign*}
\Bigg|P(&C_n(u,v)-uv \leq  x) -\Phi\left(\sqrt{\frac{n}{d}}x\right)\Bigg|= \\ &=\left|P\left(\frac{nC_n(u,v)-\lfloor nu \rfloor p_n}{\sigma_n}\leq \frac{nx+nuv-\lfloor nu \rfloor p_n}{\sigma_n}\right) - \Phi\left(\sqrt{\frac{n}{d}}x\right)\right|\leq\\
&\leq\underbrace{ \left|P\left(\frac{nC_n(u,v)-\lfloor nu \rfloor p_n}{\sigma_n}\leq \frac{nx+nuv-\lfloor nu \rfloor p_n}{\sigma_n}\right) - \Phi\left(\frac{nx+nuv-\lfloor nu \rfloor p_n}{\sigma_n}\right)\right|}_{\mathrm{(I)}} + \\
&+ \underbrace{\left| \Phi\left( \frac{nx+nuv-\lfloor nu \rfloor p_n}{\sigma_n}\right) - \Phi\left( \sqrt{\frac{n}{d}}x\right) \right|}_{\mathrm{(II)}}.
\end{flalign*}
We proceed to bound each term, starting with the first one. Here, we use inequality (2.5) introduced in \cite{LAHIRI20073570}. To simplify calculations and avoid any ambiguity, we adopt the following notation, as in the cited paper:  $p_n = \lfloor nv \rfloor /n$ and $f_n = \lfloor nu \rfloor /n.$ Moreover, let $\sigma^2_n\equiv np_n(1-p_n)f_n(1-f_n)$ and $d_n = \sigma_n^2/n.$ From the mentioned inequality
\begin{flalign}
    \mathrm{(I)} \leq \frac{C_1}{\sigma_n} \frac{1+\xi_n^2}{\lambda(\xi_n)}\exp(-C_2\xi_n^2\lambda^2(\xi_n)), \label{glowna_nierownosc}
\end{flalign}
where $\lambda(\xi_n) = (1-p_n)I(\xi_n\leq 0) + p_nI(\xi_n\geq 0)$ and $\xi_n=\xi(n,x) = (nx+nuv-\lfloor nu \rfloor p_n)/\sigma_n.$ Furthermore, the constants $C_1, C_2$ are some universal positive numbers (independent of $x$ and $n$). Notice that
\begin{align}
    \xi_n\geq0 \Leftrightarrow nx+nuv-\lfloor nu \rfloor p_n \geq 0 \Leftrightarrow x\geq \frac{\lfloor nu \rfloor\lfloor nv \rfloor}{n^2}-uv
\end{align}
and therefore 
\begin{align}
    \lambda(\xi_n) = 
    \begin{cases}
        p_n & \text{for } x > \frac{\lfloor nu \rfloor\lfloor nv \rfloor}{n^2}-uv\\
        1 & \text{for } x = \frac{\lfloor nu \rfloor\lfloor nv \rfloor}{n^2}-uv\\ 
        1-p_n & \text{for } x < \frac{\lfloor nu \rfloor\lfloor nv \rfloor}{n^2}-uv 
    \end{cases} \hspace{0.2cm}.
\end{align}
It is easy to show that 
\[
-\frac{2}{n}<-\frac{1}{n} (u+v) \leq f_np_n-uv \leq 0.
\]
For clarity in subsequent computations involving the above bounds, we introduce the following symbols
\begin{align*}
    d_- &= d_-(u,v) =  (v-1/n_0)(1-v)(u-1/n_0)(1-u), \nonumber \\
    d_+ &= d_+(u,v) =  uv,
\end{align*}
where $n_0 = \max\{n_u, n_v\}$ and $1/n_u < u \leq 1/(n_u-1).$ It is also worth noting that $d_-\leq d_n \leq d_+$ for appropriate large $n.$

We show the proof assuming that $\lambda(\xi_n) = 1-p_n.$ The same line of reasoning can be applied to the remaining two cases. Then $x<-2/n<\lfloor nu \rfloor\lfloor nv \rfloor/n^2-uv \leq 0.$ Fix any $x<0.$ From now on, the bounds below will be of asymptotic character (i.e. they hold for sufficiently large $n$, in particular $x<-2/n$). We have 
\begin{align}
    \xi_n^2 &= \frac{(nx+nuv-\lfloor nu \rfloor p_n)^2}{np_n(1-p_n)f_n(1-f_n)} \leq \frac{(nx+nuv-\lfloor nu \rfloor p_n)^2}{nd_-} =\frac{n(x+uv-p_nf_n)^2}{d_-}\leq \frac{nx^2}{d_-} \label{part_2}
\end{align}
and 
\begin{align}
    \xi_n^2 &= \frac{(nx+nuv-\lfloor nu \rfloor p_n)^2}{np_n(1-p_n)f_n(1-f_n)} \geq \frac{n(x+uv-p_nf_n)^2}{d_+}\geq \frac{n\left(x+\frac{2}{n}\right)^2}{d_+}. \label{part_3}
\end{align}
The second term of \eqref{glowna_nierownosc} can be bounded by: 
\begin{align}
    \frac{1+\xi_n^2}{\lambda(\xi_n)} \leq \frac{1}{1-v} + \frac{nx^2}{d_-(1-v)}\leq \frac{1}{d_-(1-v)} \cdot \left(1+nx^2\right) \nonumber.
\end{align}
As a result of the previous steps, we obtain the following
\begin{align}
     \mathrm{(I)} \leq \frac{C_1}{\sigma_n} \frac{1+\xi_n^2}{\lambda(\xi_n)}\exp(-C_2\xi_n^2\lambda^2(\xi_n)) 
     \leq \widehat{C}_1\left(\frac{1+nx^2}{\sqrt{n}}\right)\exp\left(-\widehat{C}_2\frac{nx^2}{2}\right),  \label{eq:pierwsza_nierownosc}
\end{align}
some positive constants $\widehat{C}_1$ and $\widehat{C}_2$, independent of $x$ and $n.$

We have established the first bound and now turn to the second inequality, which requires a slightly different approach. To show this, we apply the Mean Value Theorem to the standard normal cdf as follows 
\begin{align}
    \left|\Phi\left(\sqrt{\frac{n}{d_n}}(x+uv-p_nf_n)\right)-\Phi\left(\sqrt{\frac{n}{d_n}}x\right)\right| &\leq \sqrt{\frac{n}{d_n}}(uv-p_nf_n)\exp\left(-\frac{n(x+uv-p_nf_n)^2}{2d_n}\right) \nonumber\\
    &\leq C_3 \frac{1}{\sqrt{n}}\exp\left(-\frac{nx^2}{4}\right), \nonumber
\end{align}
for $x<0$ and some positive constant $C_3,$ independent of $x$ and $n.$ Finally, based on the Taylor expansion and triangle inequality, we have 
$$\left|\Phi\left(\sqrt{\frac{n}{d_n}}x\right)-\Phi\left(\sqrt{\frac{n}{d}}x\right)\right| = \mathcal{O}\left(\frac{1}{\sqrt{n}}\exp\left(\frac{-nx^2}{4}\right)\right),$$
Therefore, we have
\begin{align}
\mathrm{(II)} \leq C_4 \frac{1}{\sqrt{n}}\exp\left(-\frac{nx^2}{4}\right), \label{eq:druga_nierownosc}
\end{align}
where $C_4$ is some positive constant, independent of $x$ and $n$. From the inequalities \eqref{eq:pierwsza_nierownosc} and \eqref{eq:druga_nierownosc} follows 
\begin{align}
    \Bigg|P(&C_n(u,v)-uv \leq  x) -\Phi\left(\sqrt{\frac{n}{d}}x\right)\Bigg| = \mathcal{O}\left(\left(\sqrt{n}x^2+\frac{1}{\sqrt{n}}\right)\exp\left(- C_5nx^2\right)\right),
    \label{eq:koncowa_nierownosc}
\end{align}
for every $x<0$ for sufficiently large $n$ and positive constant $C_5,$ independent of $x$ and $n.$

\nocite{*}
\bibliographystyle{abbrvnat} 
\bibliography{bibliografia}

\clearpage
\appendix
\section{Supplementary materials}

\subsection{Ethanol dataset}
\begin{figure}[H]
\centering
    \includegraphics[width=1\linewidth]{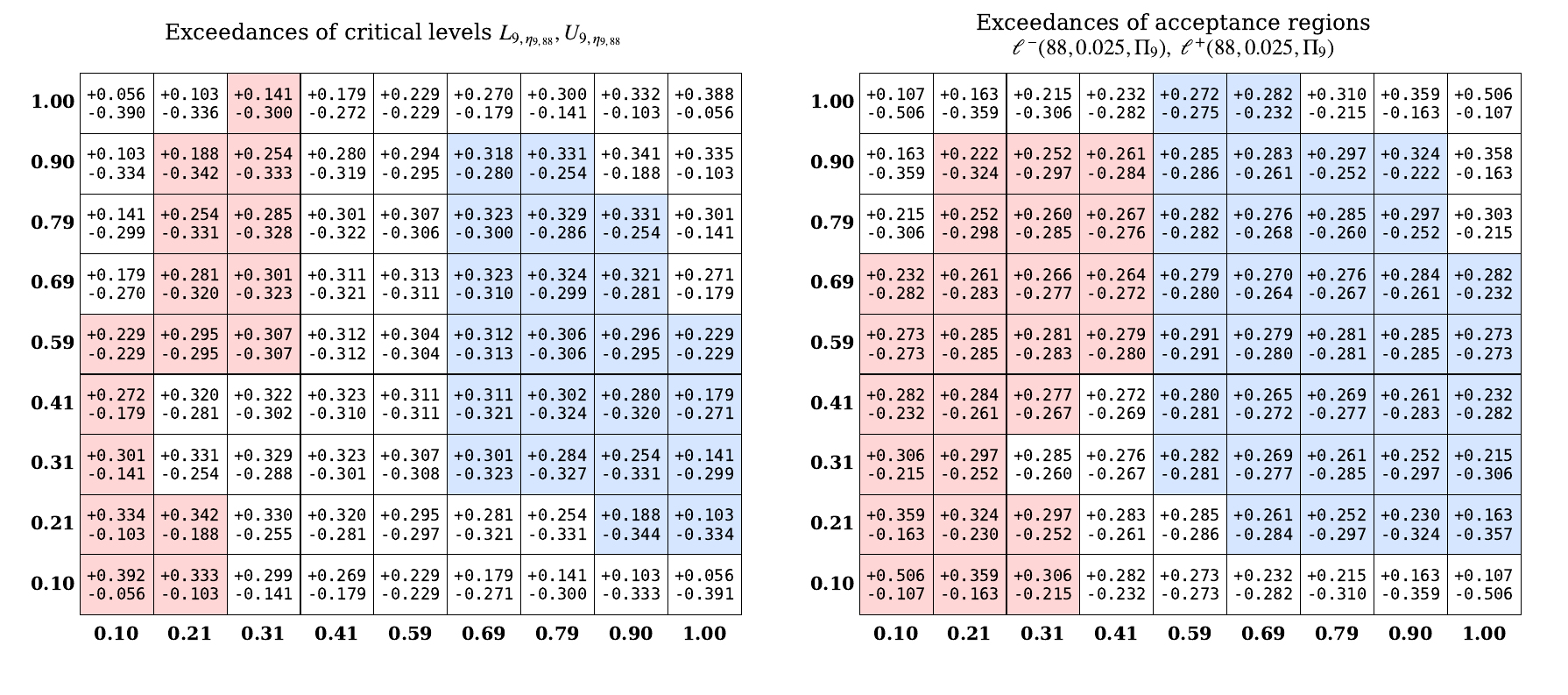}
    \caption{In each cell of the table, the first value represents the value of the upper critical surface, or upper barrier, while the second value represents the value of the lower critical surface, or lower barrier, in the corresponding region. Cells highlighted in red indicate an exceedance of the upper critical surface, whereas cells highlighted in blue indicate an exceedance of the lower critical surface.}
    \label{fig:ethanol_acc_reg}
\end{figure}

To complement the examples presented in the main text, we provide an additional example in this appendix. This example applies the same procedure described in Section 4 and is intended to further illustrate the behavior of the proposed method. The Ethanol dataset was first introduced by \cite{BrinkmanNormanD.1981EFES} and later became widely used in the statistical literature, including in \cite{SimonoffJeffreyS.2012Smis}. It is available in the \texttt{lattice} R package. The dataset contains measurements of nitrogen oxide emissions generated during the combustion of ethanol fuel in a single-cylinder engine. Our analysis focuses on the relationship between nitrogen oxide concentration and the equivalence ratio, which measures the richness of the air–ethanol fuel mixture. The same dataset was also analyzed in \cite{cmiel2024detectingdependencestructurevisualization}. In our setting, the test clearly rejects the null hypothesis of independence and correctly identifies the regions where the critical surfaces are substantially exceeded.
\begin{figure}[H]
\centering
    \includegraphics[width=0.9\linewidth]{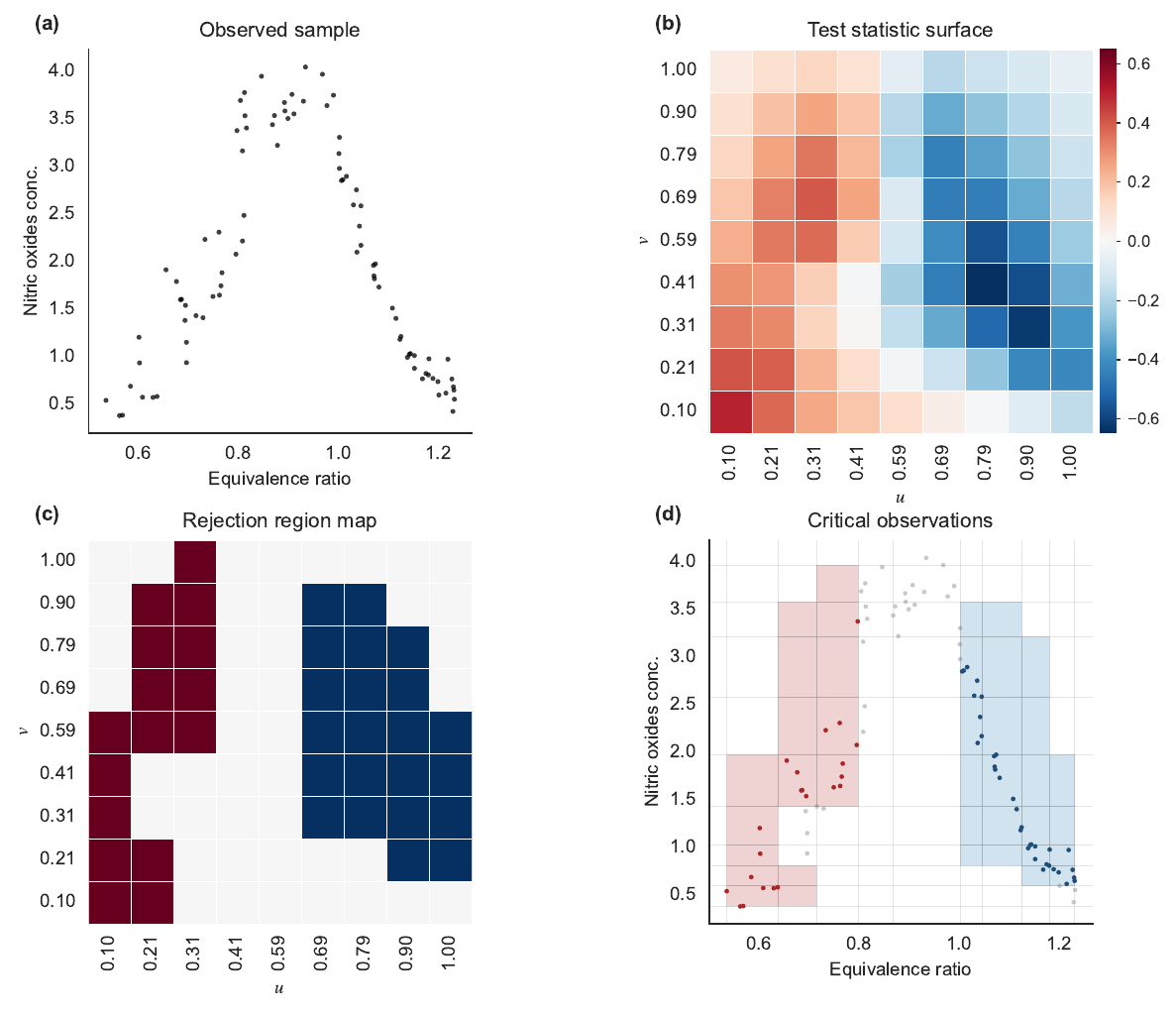}
    \caption{Panels from the upper left show: (a) observed data, with nitric oxides concentration plotted against equivalence ratio; (b) a heatmap of the smoothed estimator $q_{k,n}$; (c) a heatmap showing regions where the critical surfaces are exceeded significantly; and (d) the observed data with points highlighted as identified in the rejection map. Results are based on $10^6$ Monte Carlo simulations with $k=9$ and $\eta_{9,88}(5\%) \cong 0.1318\%.$}
    \label{fig:ethanol}
\end{figure}
The comparison between the critical surfaces and the acceptance regions proposed in \cite{cmiel2024detectingdependencestructurevisualization}, shown in Figure \ref{fig:ethanol_acc_reg}, is of particular interest. The critical surfaces identify fewer areas as significant than the acceptance-region approach. This difference arises because our procedure performs the local tests at the adjusted significance level $\eta_{k,n}(\alpha)$ which is smaller than $\alpha,$ whereas the approach proposed by \cite{cmiel2024detectingdependencestructurevisualization} tests each area separately at the nominal significance level $\alpha.$ A comparison of the boundary values shows that, near the edges, the critical-surface values are lower in absolute value than the corresponding acceptance-region values due to averaging, while the opposite pattern is observed in the central part of the square.

\subsection{Rate of \texorpdfstring{$\eta_{k,n}(\alpha)$}{eta(k,n)(alpha)} convergence}
\begin{figure}[h]
    \centering
    \includegraphics[width=0.5\linewidth]{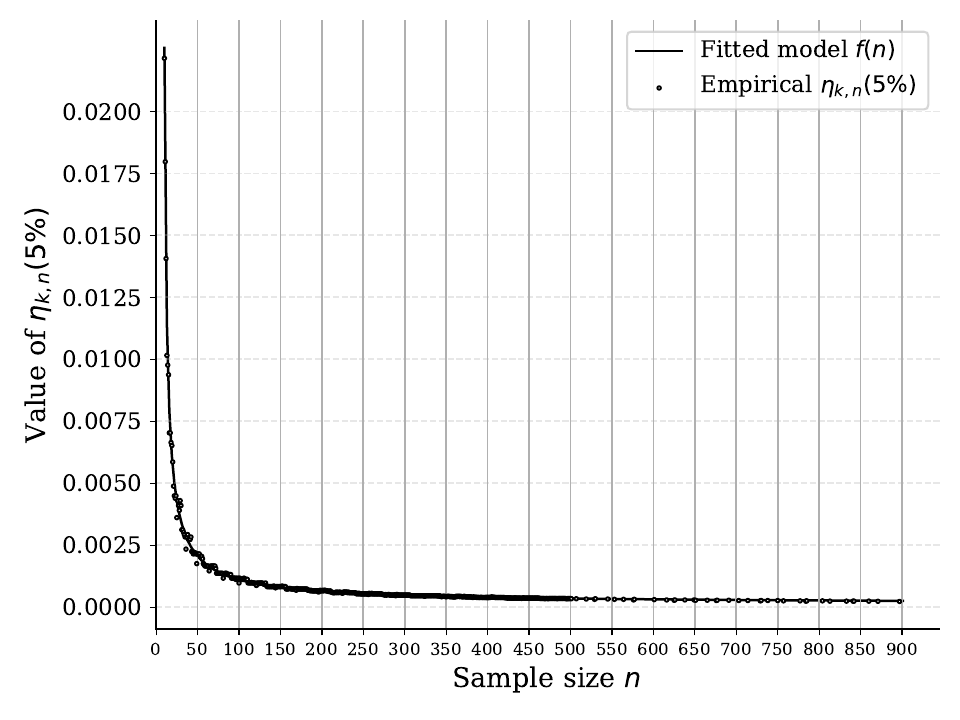}
    \caption{The $\eta_{k,n}$ values were computed for all sample sizes $n = 10,\ldots,500$ (with a larger step for $n>500$), and a polynomial model
    \(f(n) = \left(\frac{a}{n-b}\right)^c + d \)
 with parameters \(a = 0.0482\), \(b = 7.5594\), \(c = 0.8557\), and \(d = 0.0001\) was fitted. Estimated values are reported to four decimal places.}
    \label{fig:eta_plot}
\end{figure}

As established in Proposition \ref{prop1} and in the discussion following Remark \ref{rem:rate_pow}, for any fixed sample size $n \geq 2$ and significance level $\alpha$, we have
\[
\frac{\alpha}{k^2} \leq \eta_{k,n}(\alpha) \leq \alpha.
\]
A natural question concerns the rate at which $\eta_{k,n}(\alpha)$ approaches its limiting value and whether this limit is actually zero. We do not address this issue in the present paper, leaving it as a topic for future research. Preliminary Monte Carlo simulations indicate a rapid decay for small sample sizes, which may suggest behavior analogous to that of a transformed power function, as illustrated in Figure \ref{fig:eta_plot}.

\end{document}